\documentclass[prl,twocolumn,superscriptaddress,showpacs,floatfix]{revtex4-1}

\usepackage[sort&compress]{natbib}
\usepackage{graphicx, times}
\usepackage{color}
\begin{document}
\title{Deep Strong Coupling Regime of the Jaynes-Cummings model}
\author{J. Casanova}
\affiliation{Departamento de Qu\'{\i}mica F\'{\i}sica, Universidad del Pa\'{\i}s Vasco-Euskal Herriko Unibertsitatea, Apdo. 644, 48080 Bilbao, Spain}
\author{G. Romero}
\affiliation{Departamento de Qu\'{\i}mica F\'{\i}sica, Universidad del Pa\'{\i}s Vasco-Euskal Herriko Unibertsitatea, Apdo. 644, 48080 Bilbao, Spain}
\author{I. Lizuain}
\affiliation{Departamento de Qu\'{\i}mica F\'{\i}sica, Universidad del Pa\'{\i}s Vasco-Euskal Herriko Unibertsitatea, Apdo. 644, 48080 Bilbao, Spain}
\author{J. J. Garc\'{\i}a-Ripoll}
\affiliation{Instituto de F\'{\i}sica Fundamental, CSIC, Serrano 113-bis, 28006 Madrid, Spain}
\author{E. Solano}
\affiliation{Departamento de Qu\'{\i}mica F\'{\i}sica, Universidad del Pa\'{\i}s Vasco-Euskal Herriko Unibertsitatea, Apdo. 644, 48080 Bilbao, Spain}
\affiliation{IKERBASQUE, Basque Foundation for Science, Alameda Urquijo 36, 48011 Bilbao, Spain}
\date{\today}
\pacs{}
					
\begin{abstract}
We study the quantum dynamics of a two-level system interacting with a quantized harmonic oscillator in the {\it deep strong coupling  regime} (DSC) of the Jaynes-Cummings model, that is, when the coupling strength $g$ is comparable or larger than the oscillator frequency $\omega$ ($g / \omega \gtrsim 1$). In this case, the rotating-wave approximation cannot be applied or treated perturbatively in general. We propose an intuitive and predictive physical frame to describe the DSC regime where {\it photon number wavepackets} bounce back and forth along {\it parity chains} of the Hilbert space, while producing {\it collapse and revivals} of the initial population. We exemplify our physical frame with numerical and analytical considerations in the qubit population, photon statistics, and Wigner phase space.
\end{abstract}

\maketitle

The interaction between a two-level system and a harmonic oscillator is ubiquitous in different physical setups, ranging from quantum optics to condensed matter and applications to quantum information. Typically, due to the parameter accessibility of most experiments, the rotating-wave approximation (RWA) can be applied producing a solvable dynamics called the Jaynes-Cummings (JC) model~\cite{jaynescummings63}. In this case, Rabi oscillations inside the JC doublets or collapses and revivals of the system populations~\cite{Eberly80} are paradigmatic examples of the intuitive physics behind the JC dynamics. To achieve these and other phenomena in the lab, the strong coupling (SC) regime is required, that is, the qubit-oscillator coupling has to be comparable or larger than all decoherence rates. This model accurately describes the dynamics of cavity QED~\cite{raimond01,walther06}, trapped ion experiments~\cite{leibfried03}, and several setups in mesoscopic physics, where the qubit-oscillator model is essential in modeling superconducting qubits~\cite{clarke08} with either coplanar transmission lines~\cite{wallraff04,blais04,chiorescu04,hofheinz09} or nanomechanical resonators~\cite{cleland04,lahaye09}. Nowadays, solid-state semiconductor~\cite{gunter09} or superconductor systems~\cite{sornborger04,abdumalikov08,bourassa09,borja09,Lizuain10,gross10,paul10} have allowed the advent of the ultrastrong coupling (USC) regime, where the coupling strength is comparable or larger than appreciable fractions of the mode frequency: $g / \omega \gtrsim 0.1$. In this regime, the RWA breaks down and the model becomes analytically unsolvable, although some limits can be explored~\cite{irish07,chen08,hwang10,nori10,chen10}. Confident of the impressive fast development of current technology, one could explore further regimes where the rate between the coupling strength and oscillator frequency  could reach $g/\omega  \gtrsim 1$, here called deep SC (DSC) regime. This unusual regime, yet to be experimentally explored, is the focus of our current efforts. In this letter, we introduce a rigorous and intuitive description of the DSC regime of the JC model, providing an insightful picture where photon number wavepackets propagate coherently along two independent parity chains of states. In this way, the Hilbert space splits in two independent chains, exhibiting a comprehensible collapse-revival pattern of the system populations.

We consider the Jaynes-Cummings Hamiltonian without the RWA, also called Rabi Hamiltonian, describing a two-level system coupled to a single mode harmonic oscillator
\begin{equation}
 H=\frac{\hbar}{2} \omega_0 \sigma_z + \hbar \omega a^{\dag} a + \hbar g (\sigma^{+} + \sigma^{-})(a + a^{\dag}).
\label{tHamiltonian}
\end{equation} 
Here, $a$ and $a^{\dag}$ are the annihilation and creation operators of the mode with frequency $\omega,$ while $\sigma_z$ and $\sigma^{\pm}$ are Pauli operators associated to a qubit with ground state $| \rm g \rangle$, excited state $| \rm e \rangle$, and transition frequency $\omega_0$. We concentrate in the study of the DSC regime, $g / \omega \gtrsim 1$, with no particular relation between $\omega$ and $\omega_0$. We do not refer to any particular system because several of them in quantum optics and condensed matter may profit from the physical insight developed here~\cite{comment1}. We start by observing that the parity operator~\cite{shore73}
\begin{equation}
  \label{parity}
  \Pi=-\sigma_z (-1)^{n_a} = - (| {\rm e} \rangle \langle { \rm e} | - | { \rm g } \rangle \langle {\rm g} |) (-1)^{a^{\dag} a} ,
\end{equation}
with $\Pi | p \rangle = p | p \rangle$ and $p = \pm 1$, is a key element associated to the Hamiltonian in Eq.~(\ref{tHamiltonian}). It is instrumental to understand how the system dynamics moves inside the Hilbert space split in two unconnected subspaces or parity chains,
\begin{eqnarray}
 | {\rm g} 0_a \rangle \leftrightarrow | {\rm e} 1_a \rangle \leftrightarrow | {\rm g} 2_a \rangle \leftrightarrow | {\rm e} 3_a \rangle \leftrightarrow  \ldots (p = +1), \nonumber \\
 | {\rm e} 0_a \rangle \leftrightarrow | {\rm g} 1_a \rangle \leftrightarrow | {\rm e} 2_a \rangle \leftrightarrow | {\rm g} 3_a \rangle \leftrightarrow \ldots ( p = -1).
\label{chains}
\end{eqnarray}
Neighboring states within each parity chain may be connected via either rotating or counter-rotating terms. For example, in the parity chain with $p = +1$, the counter-rotating term $\sigma^{+} a^{\dagger}$ induces the transition $| {\rm g} 2_a \rangle \rightarrow | {\rm e} 3_a \rangle$, while the rotating term $\sigma^{+} a$ induces $| {\rm e} 1_a \rangle \leftarrow | {\rm g} 2_a \rangle$. When going back from DSC$\rightarrow$USC$\rightarrow$SC, the parity chains break into the known Jaynes-Cummings doublets $\{ | {\rm g} , n _a+ 1 \rangle , | {\rm e} , n_a \rangle \}$ because we enter into the domain of applicability of the RWA.

\begin{figure}[t]
    \includegraphics[width=\linewidth]{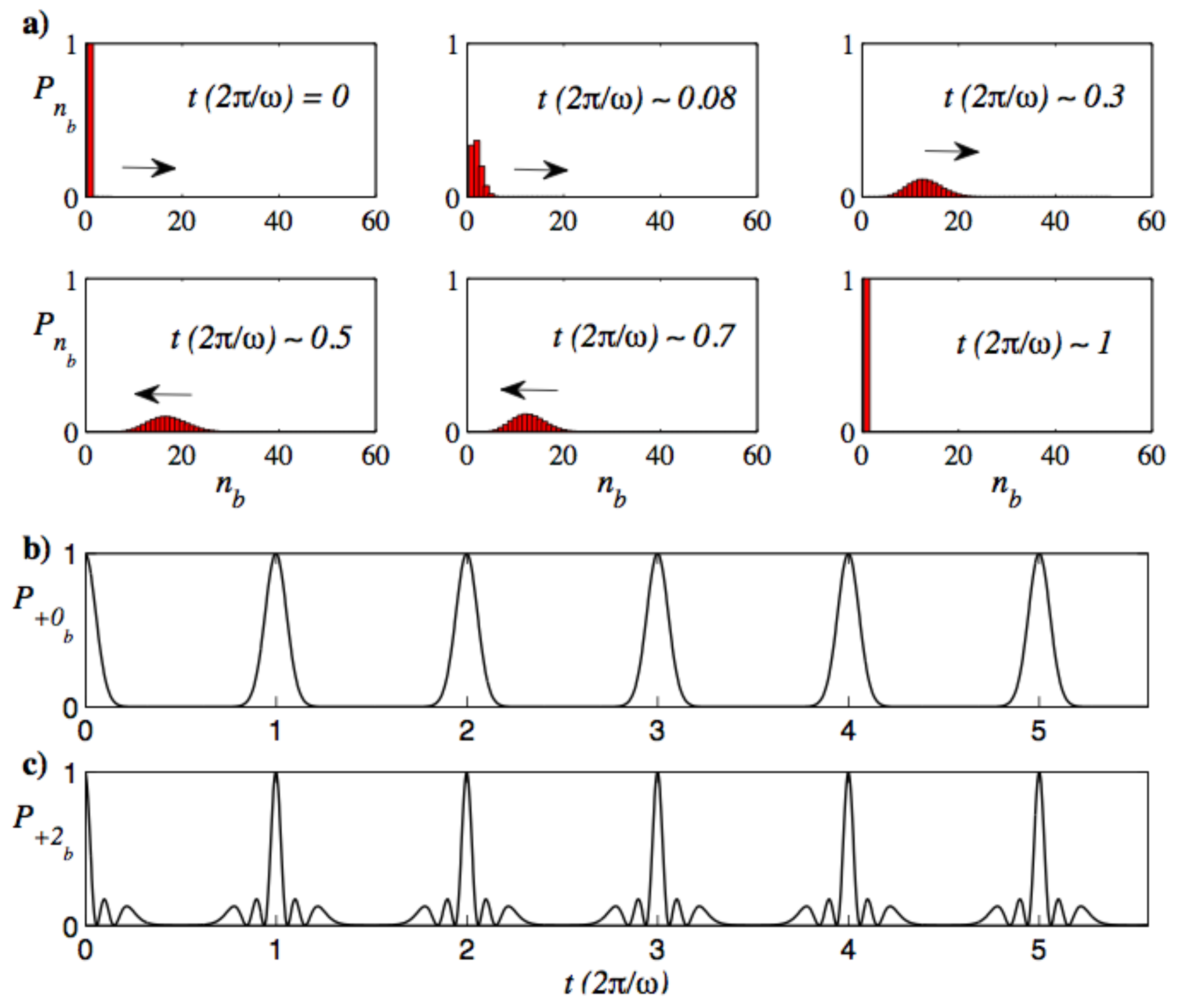}
  \caption{(Color online) (a)-(b) Round trip of a photon number wavepacket and collapse-revivals due to DSC dynamics with initial state $| + ,  0_b \rangle = | {\rm g} ,  0_a \rangle$. c) Collapse-revivals with secondary peaks due to counterpropagating photon number wavepackets starting in initial state $| + ,  2_b \rangle = | {\rm g} ,  2_a \rangle$. For all cases, $\omega_0 = 0$ and $g / \omega = 2$. }
  \label{Fig1}
\end{figure}

We introduce the parity basis $| p ,n_b\rangle,$ where $b^{\dagger}b | n_b \rangle = n_b | n_b \rangle$, and $b = \sigma_x a$ such that $\, b | p ,n_b \rangle = \sqrt{n_b} | p ,n_b - 1 \rangle$.
Using this basis, the Hamiltonian in Eq.~(\ref{tHamiltonian}) can be written as
\begin{equation}
H = \hbar \omega b^{\dag} b+\hbar g (b+b^{\dag})-\hbar\frac{\omega_0}{2}(-1)^{b^{\dag}b}\Pi .
\label{pHamiltonian}
\end{equation}
This Hamiltonian commutes with the parity operator $\Pi$, and for each parity chain ($p = \pm 1$) there is an independent Hamiltonian describing a perturbed harmonic oscillator. Note that the term $-\hbar \omega_0(-1)^{b^{\dag}b}\Pi/2$ behaves as an energy shift proportional to $\omega_0$. In the DSC regime, we can get rid of the term $\hbar g (b+b^\dagger)$ in Eq.~(\ref{pHamiltonian}) by changing to the basis $D(-\beta_0)|p , n_b \rangle$, with $D(\beta_0)=e^{\beta_0 b^\dagger - \beta_0^*b}$ and $\beta_0=g/\omega$. The eigenenergies and eigenfunctions can be approximated as a series in $\omega_0/\omega$
\begin{eqnarray}
  E^{\beta_0}_{p,n_b} /\hbar &\approx& \omega n_b  - g^2/\omega -
  \frac{\omega_0}2 p (-1)^{n_b} \Delta_{n_b n_b} + \label{eq:approx}\\
  &+&\sum_{m_b \neq n_b} \frac{\omega_0^2}{4\omega(n_b - m_b)}|\Delta_{n_b m_b}|^2 + \mathcal{O}(\omega_0^3/\omega^3).\nonumber
\end{eqnarray}
Alternative approximations can be found in the literature~\cite{irish07}. To first order we get a displacement in the energy levels due to the coupling $\Delta_{n_b n_b} = \langle n_b | D(2\beta_0)| n_b \rangle,$ a correction which is much smaller than one, $|\Delta_{n_b m_b}| \ll 2^{-(n_b + m_b)}.$  Note that this formalism is rigorously valid in the DSC regime.

\begin{figure}[t]
   \includegraphics[width=8cm]{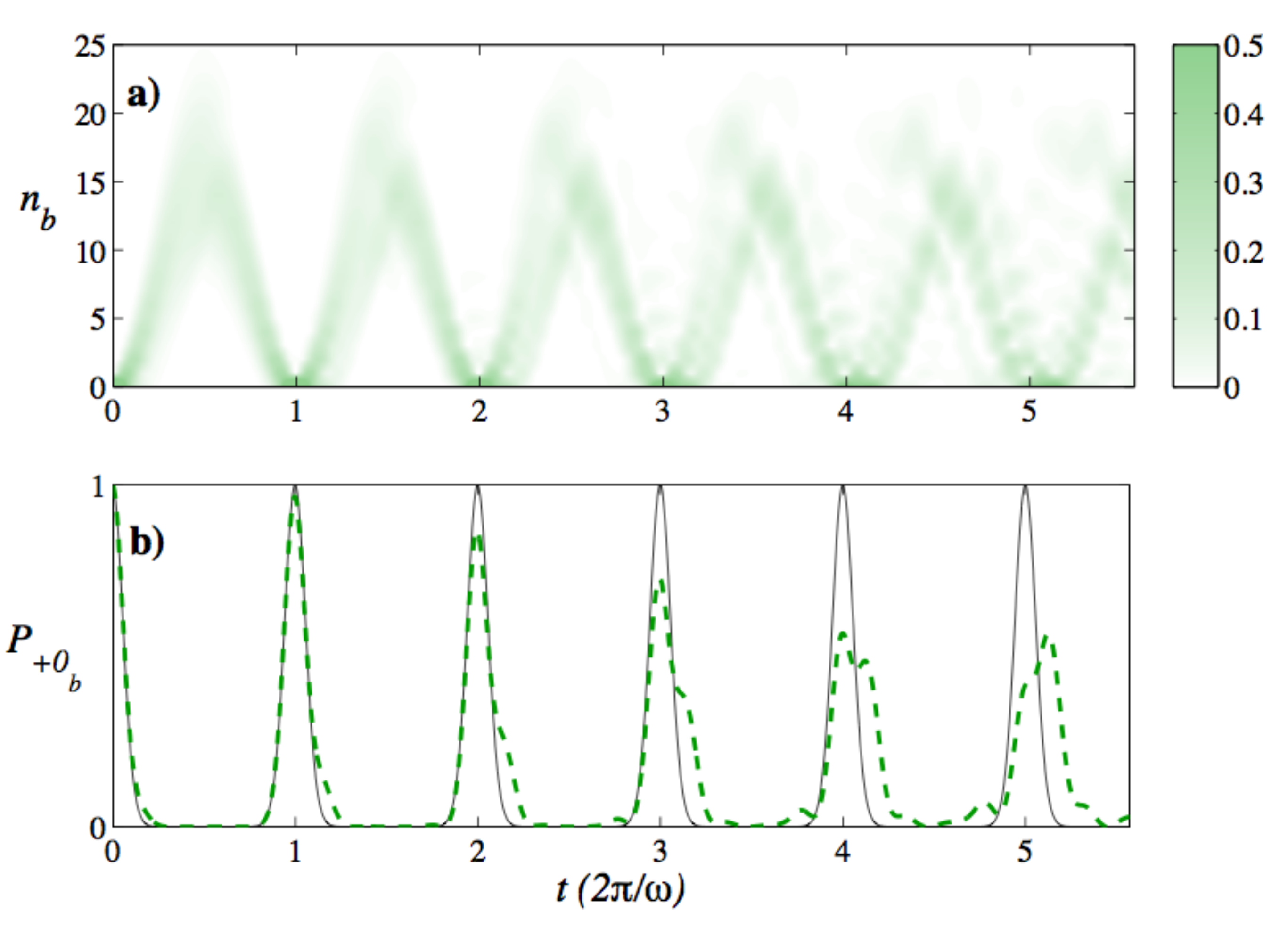}
   \caption{(Color online) (a) Photon statistics at different times of the evolution with $\omega_0=0.5~\omega$. (b) Comparison of probability $P_{+,0_b}(t)$ calculated for $\omega_0=0$ (solid line) and $\omega_0=0.5~\omega$ (dashed line). In all simulations the initial state is $| + , 0_b \rangle$ and $g/\omega=2$.}
   \label{Fig2}
\end{figure}

We study now the DSC dynamics with the initial state $|\psi(0)\rangle = | + , 0_b \rangle=| {\rm g} , 0_a \rangle$, as we activate the interaction in Eq.~(\ref{pHamiltonian}). We observe that the photon statistics, $P_{n_b} (t)$, will spread independently along each parity chain, eventually reaching an energy barrier and bouncing repeatedly.  Remarkably, an intuitive picture can be found, as displayed in Figs.~\ref{Fig1} and \ref{Fig2} , that provides physical insight in a problem that is, in general, analytically intractable. Note that, in Figs.~\ref{Fig1}a and \ref{Fig1}b, the round trip of the initial photon number wave packet induces collapse-revivals that are not reminiscent of the SC regime of the JC model~\cite{Eberly80}, where initial large coherent states are required. In the DSC limit, with $\omega_0 = 0$, this intuitive picture can be rigorously confirmed integrating the evolution
\begin{eqnarray}
  |\psi(t)\rangle &=& D^{\dagger} (\beta_0)e^{-i(\omega b^\dagger b - g^2/\omega) t} D(\beta_0)
  | + , 0_b \rangle \label{eq:U0}\\
  &=& U(t,\omega_0=0)|\psi(0)\rangle = e^{i \frac{g^2}{\omega} t} e^{- i (\frac{g}{\omega})^2 \sin(\omega t)}  | + , \beta(t) \rangle , \nonumber
\end{eqnarray}
where $\beta(t)=\beta_0(e^{-i\omega t}-1)$ is the amplitude of a coherent state. The revival probability of the initial state reads
\begin{equation}
  P_{+0_b}(t) = |\langle \psi(0)|\psi(t)\rangle|^2 = e^{-|\beta(t)|^2} ,
\end{equation}
exhibiting periodic collapses and full revivals~\cite{comment2}. When the initial state is $| + ,  2_b \rangle = | {\rm g} ,  2_a \rangle$, as in Fig.~\ref{Fig1}c, the DSC dynamics generates counterpropagating photon number wavepackets in both directions that bounce back and forth producing interference secondary peaks. Similar intuition follows when considering initial superposition states, e. g. $( | + ,  0_b \rangle + | + ,  2_b \rangle) / \sqrt{2}$, as long as the state components belong to the same parity chain, otherwise no secondary peaks appear. When we break the qubit degeneracy, $\omega_0 \ne 0,$ the intuitive picture remains but we lose the integrability of the problem. Probability still spreads along each parity chain, as seen in Fig.~\ref{Fig2}, but now the photon number wavepacket suffers self-interference, it distorts and its center no longer follows the periodic orbits of $\omega_0=0.$ The result are full collapses and partial revivals where probability $P_{+0_b}$ is not completely restored, and whose maximum value deteriorate as time passes.

\begin{figure}[t]
  \includegraphics[width=7.5cm]{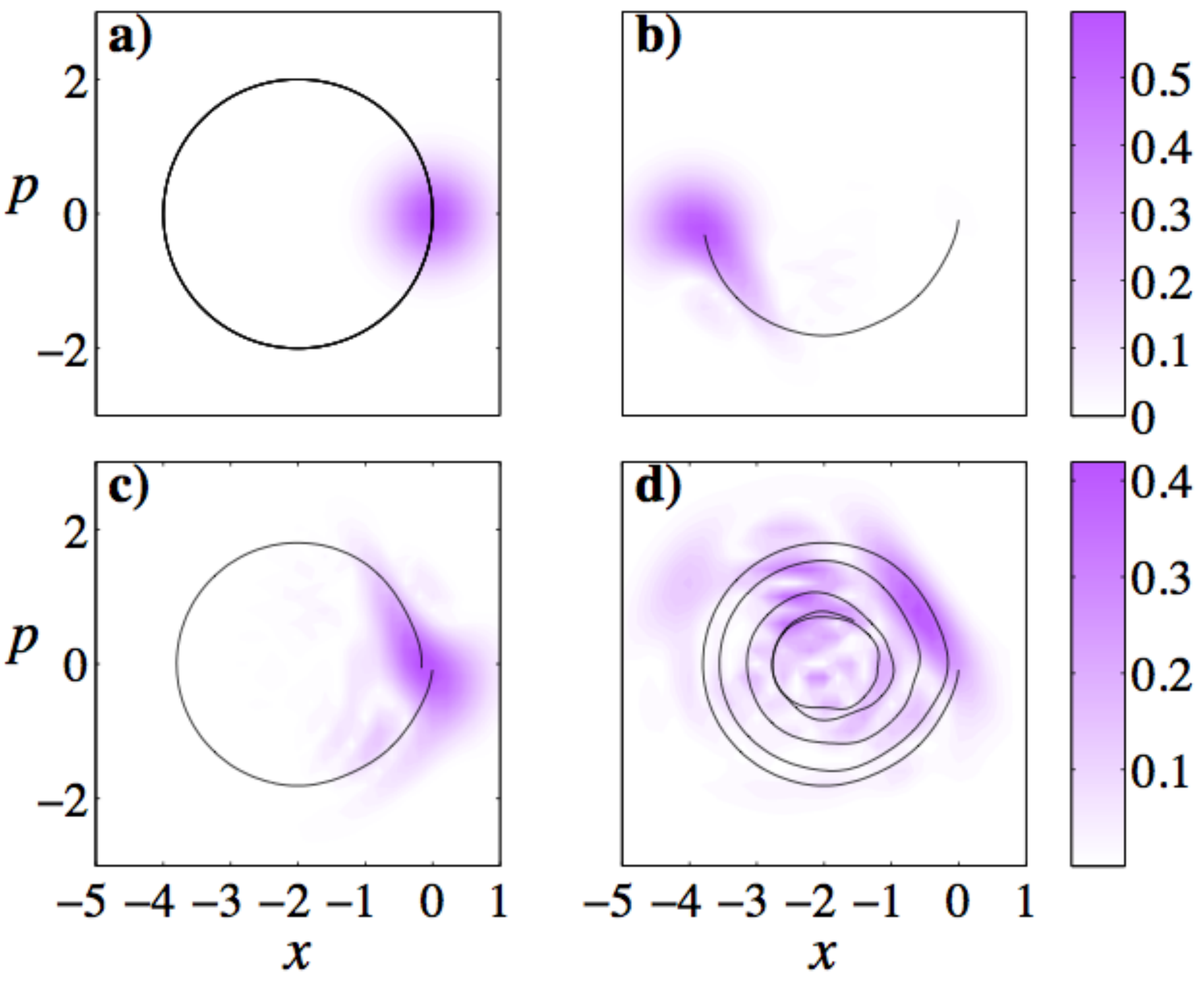}
  \caption{(Color online) Modulus of the Wigner function, $|W(x,p)|,$ and trajectory $(x , p),$ for Eq.~(\ref{pHamiltonian}). (a) For $\omega_0=0$ the Wigner function is a symmetric Gaussian (a coherent state) and moves clockwise in phase space along a circle. (b)-(c)-(d) For $\omega_0=0.5~\omega$ the Wigner function loses its symmetry and moves clockwise along a spiral trajectory as shown in the sequence corresponding to $t = 0.5, 1, 5$. In both simulations the initial state is $| + , 0_b \rangle$ and $g/\omega=2$.}
  \label{Fig3}
\end{figure}
The collapses and revivals have also interesting consequences in phase space, which we have analyzed using the Wigner function and phase space trajectories, $(\bar{x}(t),\bar{p}(t))=\langle (b + b^\dagger, i b^\dagger - i b)/\sqrt{2}\rangle.$ In the integrable case, $\omega_0=0$, the Wigner function of the state is a Gaussian centered on a point $x+ip=\beta(t)$ which draws periodic circular orbits on the plane. As soon as we switch on the term proportional to $\omega_0,$ the wavepacket suffers two distortions, see Fig.~\ref{Fig3}. The first one is a squeezing tangential to the orbit, shown in Fig.~\ref{Fig3}c. Accompanied by the difussion and interference in the Wigner wavepacket, the orbits also distort, becoming spirals that relax towards the center of the original orbits, $-\beta_0.$

The phenomenon of collapses and revivals for the nonintegrable case, $\omega_0 \neq 0$, even if partial, reveal a structure in the Hamiltonian spectrum, which is approximately equispaced. We can write the revival probability
\begin{equation}
  P_{+0_b}(t) = {\Large|} \sum_{\ell} | \langle \psi(0)|\phi_{\ell} \rangle |^2 e^{-i E_{\ell} t/\hbar} {\Large|}^2,
  \label{eq:revivals}
\end{equation}
as a function of the overlap of the initial state with the eigenstates of the full model, $(H - E_{\ell})| \phi_{\ell} \rangle=0.$ When $\omega_0 = 0,$ the eigenenergies are regularly spaced, $E_{\ell} = \hbar \omega \ell$ and the function becomes periodic with period $2\pi/\omega.$ This causes an initial Gaussian wavepacket in phase space to get reconstructed at the same position for $t=2\pi,4\pi,6\pi,\ldots.$ In the DSC case with $\omega_0\neq 0,$ the energy levels deviate very slightly from this regular distribution, $E_{\ell} = \hbar\omega {\ell} - \hbar \delta_{\ell} \omega_0,$ where the correction $\delta_{\ell} \omega_0$ is less than $10\%$ for the examples considered in this work. The reconstruction of the wavepackets is incomplete and different partial waves may get delayed or accelerated with respect to the original orbit, $\beta(t)$. This causes the squeezing of the Wigner function and the self-interference in the photon number wavepacket, as displayed in Fig.~\ref{Fig2}a.

\begin{figure}[t]
  \includegraphics[width=7.5cm]{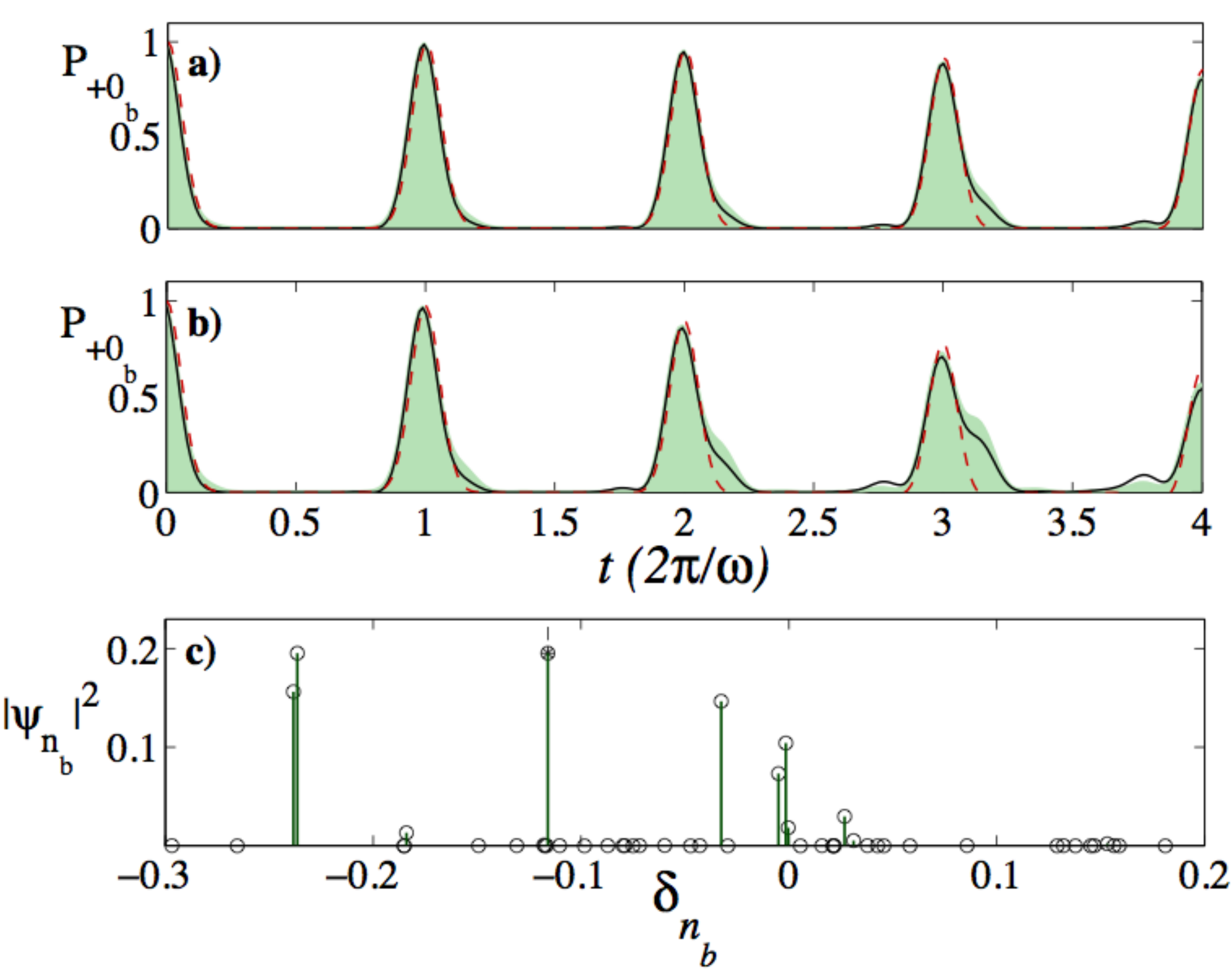}
  \caption{(Color online) Collapses and revivals of $P_{+,0_b}(t)$ for $g/\omega=2$ and (a) $\omega_0=0.3\omega$ and (b) $\omega_0=0.5\omega.$ We plot the exact numerical solution (area) at first order in $\omega_0/\omega$ (solid) and a two-mode approximation (red, dashed). (c) Distribution of probability of the different detunings $\delta_{n_b} = (\hbar\omega n_b - E_{n_b}) / \hbar\omega_0,$ weighted by their contribution to the wavefuncion given in Eq.~(\ref{eq:revivals}). We have marked the
level $N^r_b$ which is used for the two-mode approximation in curves (a) and (b).}
   \label{Fig4}
\end{figure}

We have also found that the overall dynamics is very accurately captured by the first order correction to the eigenenergies, shown in Eq.~(\ref{eq:approx}). If we use $\delta_{n_b} = (-1)^{n_b} p \Delta_{n_b n_b}/2$ in Eq.~(\ref{eq:revivals}), together with the initial condition $\langle \psi(0) | \phi_{n_b}\rangle = \langle \psi(0) | D(-\beta_0) |p {n_b} \rangle = \delta_{+,p} \exp(-|\beta|^2/2)\beta^{n_b}/{n_b}!$, we obtain curves that approximate very well the exact result. This is shown in Figs.~\ref{Fig4}a~and~\ref{Fig4}b for values of $\omega_0=0.3\omega$ and $0.5\omega$ in the DSC regime, $g/\omega=2$. In both cases the revivals happen close to $t = k \,(2\pi / \omega)$, with integer $k$, but decreasing in intensity and with a large fraction of the curve moving with a slower speed, reconstructing itself at later times. The effect of this is more evident in Fig.~\ref{Fig4}b, and also in the photon number wavepacket plot in Fig.~\ref{Fig3}a, where one appreciates two waves with slightly different periods interfering with each other. The existence of this delayed revivals is due to the structure of the wavefunction $\psi(t)$ that, as shown in Fig.~\ref{Fig4}c, is composed of many contributions close to zero detuning, $\delta_{n_b}\simeq 0,$ and a few large contributions with $\delta=-0.116,-0.223.$ The former constitute the main revivals, while the second ones make the revivals at slightly longer periods, $2\pi\omega-0.223\omega_0,$ forming the second wavefront in Fig.~\ref{Fig2}a and Fig.~\ref{Fig4}a-b.

Based on our previous results, we have developed a heuristic approximation that allows us to reproduce the main revivals. Our method recognizes that if we start the dynamics with state $|\psi(0)\rangle=|+,N_b\rangle,$ the main contribution to the wavefunction is around a level $N^r_b=[(g/\omega)^2+N_b],$ where $[\cdot]$ denotes the closest integer. This is indeed the case in the considered examples, as shown in Fig.~\ref{Fig4}c. We will only consider the energy correction for this level, $\delta_{N^r_b},$ and neglect the dephasing of all other \textit{off-resonant} terms. Under this criteria, we approximate the system state, up to normalization, as 
\begin{eqnarray}
 | \psi(t) \rangle \approx && U(t,\omega_0=0) | \psi(0) \rangle + \psi_{N^r_b}(e^{-i(\omega N_b \nonumber  -  \omega_0\Delta_{N^r_b N^r_b} /2)t} \\ && - e^{-i\omega N_b t}) D(-\beta_0) |+,N^r_b\rangle,
 \label{jcstate1} 
\end{eqnarray}
where $U(t,\omega_0=0)$ is the evolution operator at $\omega_0=0$ introduced in Eq.~(\ref{eq:U0}), and $\psi_{N^r_b}=\langle \phi_{N^r_b}|\psi(0)\rangle$. This form of the state is motivated by the behavior of the Wigner function, as seen in Fig.~\ref{Fig3}(b): a central core approximated by the solution of $\omega_0=0$ plus a delayed correction capturing the effects of $\omega_0,$ forming the squeezed tail. Considering the simplest case of $N_b = 0$, from state in Eq.~(\ref{jcstate1}), it is straightforward to obtain a simple analytical expression for the revival probability
\begin{eqnarray}\label{prob}
P_{+0_b}(t)
\approx & 2 & e^{-|\beta(t)|^2/2-\beta_0^2} \frac{\beta_0^{2N^r_b}}{N^r_b!}\left[ \cos (\omega_0 \delta_{N^r_b} t/2) -1 \right] \nonumber\\
 & + & e^{-|\beta(t)|^2},
\label{jcapprox}
\end{eqnarray} 
where small terms are neglected. This expression has been compared with the exact solution, as shown in Fig.~\ref{Fig4}a and \ref{Fig4}b, giving a good estimate of the height of the partial revivals as a function of time and $\omega_0.$ The other features, such as the delayed front is not reproduced because this approximation does not contain the contributions with $\delta=-0.223,$ but this can be improved by including more resonant levels.

\begin{figure}[t]
  \includegraphics[width=7.5cm]{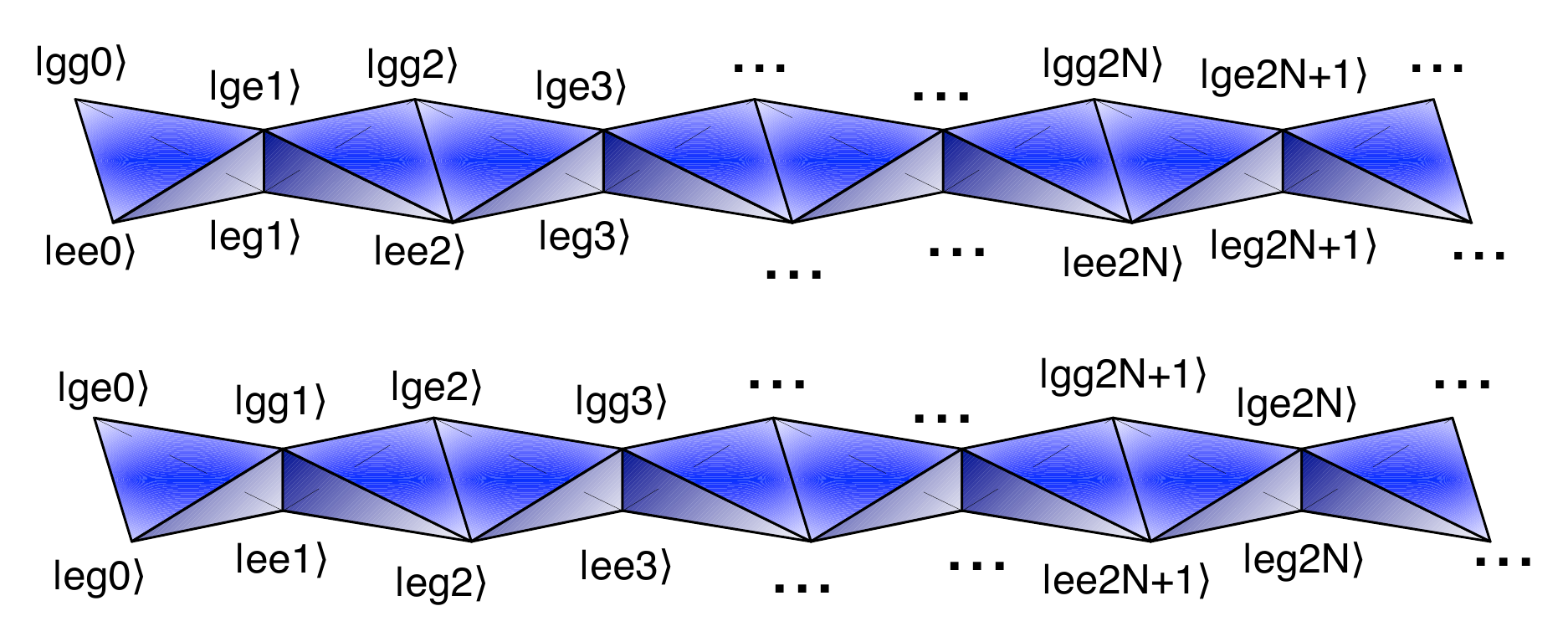}
  \caption{(Color online) Parity chains for the two-qubit DSC regime.}
   \label{Fig5}
\end{figure}   

{\em Two qubits and a mode}$.-$ It is also possible to give a qualitative and quantitative description of the DSC considering the case of two qubits. Here, the $2\otimes 2 \otimes N$-dimensional Hilbert space bifurcates into two independent parity chains of tetrahedra, see Fig.~\ref{Fig5}, where each vertex is connected to their neighbors via rotating or counter-rotating terms. The same dynamical properties of probability collapses and revivals can be found, as well as interesting entanglement properties.

{\em Conclusions}$.-$
The SC regime of the JC model is considered nowadays an intuitive and comprehensive field. The USC regime is described by the SC regime plus RWA and higher-order corrections. In this work, we have aimed at developing an insightful description of the DSC regime of the JC model. The transition between the SC and DSC regimes remains a rather diffuse crossover with well understood frontiers.

We acknowledge funding from Basque Government grants BFI08.211 and IT472-10; Spanish MICINN projects FIS2009-12773-C02-01, FIS2009-10061, and Juan de la Cierva Program; QUITEMAD, and SOLID European project.

\end{document}